\documentclass[showpacs,aps,prapplied,twocolumn,superscriptaddress,longbibliography]{revtex4-1}
\usepackage{graphicx}
\usepackage{bm}
\usepackage{color}
\usepackage{amsmath}
\usepackage{amssymb}
\usepackage{enumerate}
\usepackage{xspace}
\usepackage{mathrsfs}
\usepackage{mathptmx}
\usepackage{graphicx}
\usepackage{amsmath,amssymb}
\newcommand{\new}[1]{\textcolor{black}{#1}}

\newcommand{\Neel}{N\'eel }
\usepackage[normalem]{ulem}

\begin{document}

\newcommand{\Ham}{\ensuremath{\mathscr{H}}\xspace}
\newcommand{\ku}{\ensuremath{k_{\mathrm{u}}}\xspace}
\newcommand{\Sz}{\ensuremath{S_z}}
\newcommand{\Jij}{\ensuremath{J_{ij}}\xspace}
\newcommand{\mus}{\ensuremath{\mu_{\mathrm{s}}}\xspace}
\newcommand{\sH}{\ensuremath{\mathbf{H}}\xspace}
\newcommand{\Happ}{\ensuremath{\mathbf{H}_{\mathrm{app}}}\xspace}
\newcommand{\sS}{\ensuremath{\mathbf{S}}\xspace}
\newcommand{\TC}{\ensuremath{T_{\mathrm{C}}}\xspace}
\newcommand{\TN}{\ensuremath{T_{\mathrm{N}}}\xspace}
\newcommand{\kB}{\ensuremath{k_{\mathrm{B}}}\xspace}
\newcommand{\muB}{\ensuremath{\mu_{\mathrm{B}}}\xspace}
\newcommand{\Ms}{\ensuremath{M_{\mathrm{s}}}\xspace}

\newcommand{\vampire}{\textsc{vampire}\xspace}
\newcommand{\saf}{\textsc{saf}\xspace}
\newcommand{\abinitio}{\textit{ab-initio}\xspace}
\newcommand{\AFM}{\textsc{afm}\xspace}
\newcommand{\FCC}{\textsc{fcc}\xspace}
\newcommand{\etal}{\textit{et al}\xspace}

\newcommand{\kN}{\ensuremath{k_{ij}}\xspace}
\newcommand{\eij}{\ensuremath{\mathbf{e}_{ij}}\xspace}
\newcommand{\Jnn}{\ensuremath{J_{ij}^{\mathrm{nn}}}\xspace}
\newcommand{\Jnnn}{\ensuremath{J_{ij}^{\mathrm{nnn}}}\xspace}

\newcommand{\IrMn}{\ensuremath{\text{IrMn}_3}\xspace}
\newcommand{\MnAu}{\ensuremath{\text{Mn}_2\text{Au}}\xspace}
\newcommand{\Lonetwo}{\ensuremath{\text{L}1_2}\xspace}

\title{Magnetic stray fields in nanoscale magnetic tunnel junctions}
\author{Sarah Jenkins}
\email{sarah.jenkins@york.ac.uk}
\affiliation{Department of Physics, The University of York, York, YO10 5DD, UK}
\author{Andrea Meo}
\affiliation{Department of Physics, Mahasarakham University, Thailand}
\author{Luke E. Elliott}
\author{Stephan K. Piotrowski}
\author{Mukund Bapna}
\affiliation{Department of Physics, Carnegie Mellon University, Pittsburgh, PA, USA}
\author{Roy W. Chantrell}
\affiliation{Department of Physics, The University of York, York, YO10 5DD, UK}
\author{Sara A. Majetich}
\affiliation{Department of Physics, Carnegie Mellon University, Pittsburgh, PA, USA}
\author{Richard F. L. Evans}
\email{richard.evans@york.ac.uk}
\affiliation{Department of Physics, The University of York, York, YO10 5DD, UK}

\begin{abstract}
The magnetic stray field is an unavoidable consequence of ferromagnetic devices and sensors leading to a natural asymmetry in magnetic properties. Such asymmetry is particularly undesirable for magnetic random access memory applications where the free layer can exhibit bias. Using atomistic dipole-dipole calculations we numerically simulate the stray magnetic field emanating from the magnetic layers of a magnetic memory device with different geometries. We find that edge effects dominate the overall stray magnetic field in patterned devices and that a conventional synthetic antiferromagnet structure is only partially able to compensate the field at the free layer position. A granular reference layer is seen to provide near-field flux closure while additional patterning defects add significant complexity to the stray field in nanoscale devices. Finally we find that the stray field from a nanoscale antiferromagnet is surprisingly non-zero arising from the imperfect cancellation of magnetic sublattices due to edge defects. Our findings provide an outline of the role of different layer structures and defects in the effective stray magnetic field in nanoscale magnetic random access memory devices and atomistic calculations provide a useful tools to study the stray field effects arising from a wide range of defects. 
\end{abstract}

\maketitle

\section{Introduction} 
Magnetic Random Access Memory (MRAM) is a promising technology for low-power non-volatile device memory \cite{KhvalkovskiyJPhysD2013}. With the breakthrough of a suitable materials system in CoFeB/MgO for spin-transfer-torque MRAM (STT-MRAM) devices \cite{IkedaNatMat2010} significant progress has been made towards full-scale commercialisation and a move to non-volatile memory technology \cite{KentNatNano2015,KhvalkovskiyJPhysD2013}. A key requirement for wide-scale use of STT-MRAM is device reliability, requiring effectively unlimited write operations, but also data retention for at least 10 years and consistency of operation. Despite its apparent simplicity, the properties of ultrathin CoFeB/MgO films are surprisingly complex, with intricate magnetic interactions \cite{SatoPRB2018,BoseJPCM2016,CuadradoPRApplied2018} and nanoscale structural \cite{SankaranIEEE2018}, thermal \cite{MeoSciRep2017,BapnaAPL2016,IgarashiAPL2017} and dynamic effects\cite{Sampan-a-paiPRApplied2019}. 

One problem not often considered is that of stray magnetic fields originating from MRAM devices and also affecting their magnetic characteristics. These stray magnetic fields are a source of non-uniformity in nanoscale devices and can have a significant influence on the magnetic properties, thermal stability and switching characteristics \cite{IkedaNatMat2010}. In the supplementary information \cite{supp} we present experimental measurements of the role of these stray fields on the magnetoresistance and relaxation time of individual MTJs. The methodology is described in detail in Bapna \textit{et al} \cite{BapnaAPL2016}. The experimental data show the importance of edge fields in nanoscale devices and how these can be compensated for with different device structures.

Previous theoretical studies of magnetostatic stray fields \cite{BapnaAPL2016,DevolderJPhysD2019} have considered a continuum micromagnetic approach which is sufficient for continuous materials. However, nanoscale MTJs are only a few atoms thick and their fabrication and patterning leads to a diverse range of defects. Modelling these defects goes beyond the capabilities of micromagnetic approaches, and atomistic models are needed \cite{EvansJPCM2014}. A similar problem arises when considering the magnetostatic stray field, where the sources are no longer a uniform continuum of atoms and have inherent structural and magnetic order. This problem grows with higher temperatures where thermal spin fluctuations are significant and the dipole fields can statistically vary in time. Crucially the temperature dependence of the magnetization and finite size effects are important when considering stray fields emanating from nanoscale magnetic dots. Antiferromagnets also play an essential stabilizing role in many spintronic devices, and macroscopically their stray field is zero. At the nanoscale this is not necessarily the case and such effects are not accessible using a standard continuum magnetostatic approach.

The magnetostatic stray field for a perpendicular CoFeB/MgO/CoFeB MTJ increases as the diameter is reduced. Failure to offset the resulting loop shift causes the critical current to be larger than necessary, leading to greater power consumption. However, it is yet unclear how best to minimize this stray field. The simulations described here examine several different strategies using an atomistic dipole-dipole approach. We find that edge effects are particularly important for nanoscale MRAM devices and that defects and antiferromagnets can contribute statistical variations in the stray field leading to an additional natural variance in device properties. 

\section{Stack structures}
\begin{figure}[tb]
\centering
\includegraphics[width=8.5cm]{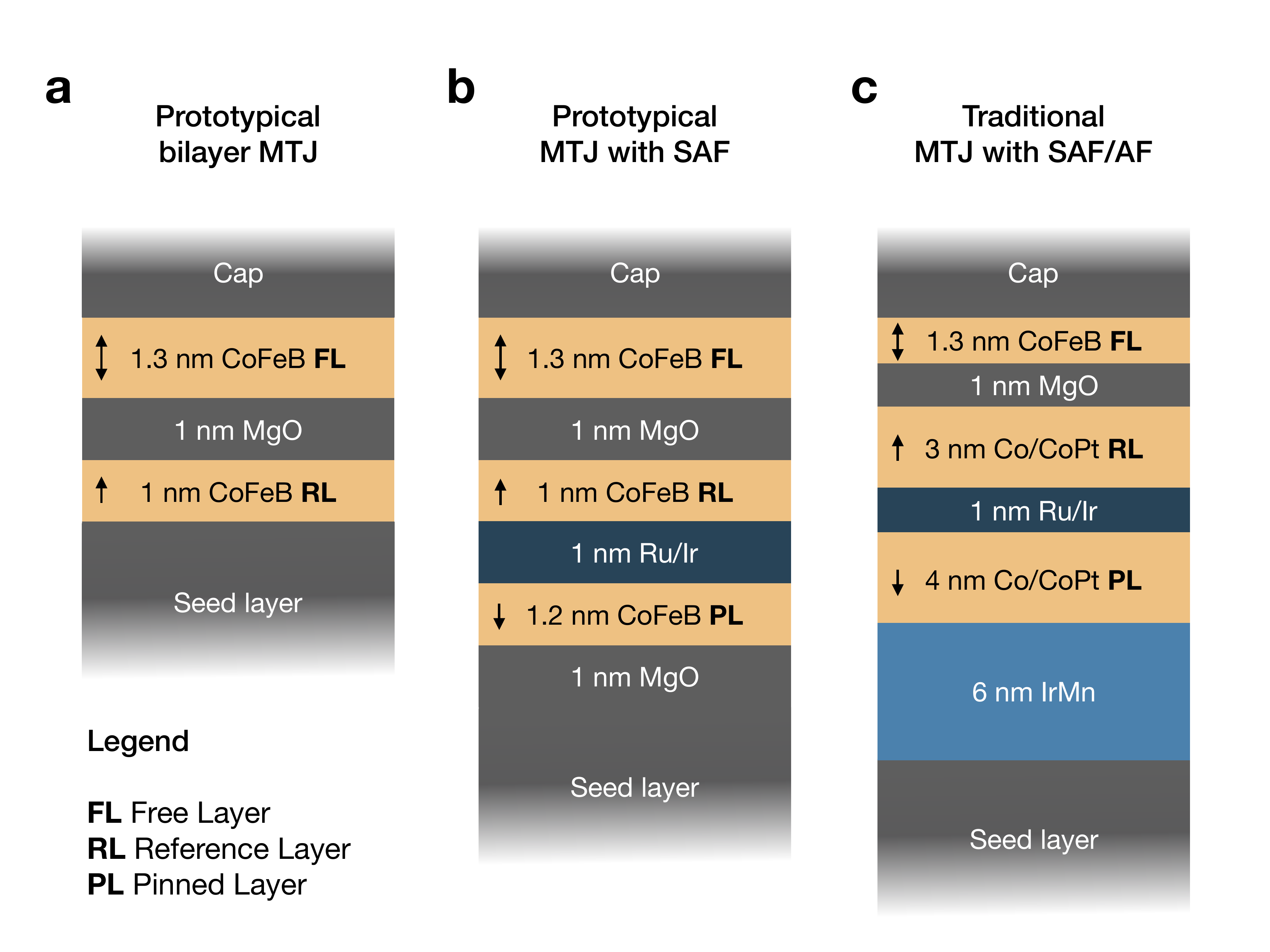}
\caption{Visualisation of alternative thin film stack structures for prototypical MTJs and practical MRAM devices for prototypical bilayer MTJ (a), prototypical bilayer MTJ with synthetic antiferromagnet reference and pinned layer(b) and more 'traditional' MTJ structure with synthetic antiferromagnet reference and pinned layers with antiferromagnet exchange biasing layer (c). Colour Online.}
\label{fig:stacks}
\end{figure}

Practical MRAM devices have a number of limitations compared to simple functioning magnetic tunnel junctions, where the devices must have high durability, high thermal stability, consistent performance, fabricatable with low annealing temperatures and manufacturable at gigabit volumes. The prototypical MTJ (Fig.~\ref{fig:stacks}(a)) satisfying the basic requirements of spin-transfer torque magnetic random access memory (STT-MRAM) consists of a bilayer of CoFeB sandwiching a thin MgO tunnel barrier\cite{IkedaNatMat2010}. The MgO layer performs two essential functions: a spin tunnelling barrier with large tunnelling magnetoresistance \cite{KhvalkovskiyJPhysD2013} and a large interfacial perpendicular magnetic anisotropy\cite{IkedaNatMat2010,KhvalkovskiyJPhysD2013,BoseJPCM2016}. The high magnetic anisotropy is essential to stabilize the magnetic orientation of the CoFeB layers and its interfacial nature gives a strong thickness dependence of the anisotropy. Therefore different thickness layers have different coercivities and threshold currents for STT switching, providing a natural reference layer (RL) and free layer (FL). The free layer is required to have lower stability than the reference layer and in simple CoFeB/MgO/CoFeB devices with a dual layer structure must be around 1.3 nm thick to ensure perpendicular anisotropy  but not too high to prevent switching\cite{IkedaNatMat2010}. Thicker layers are possible by using an additional MgO capping layer\cite{ApalkovIEEE2016,DevolderJPhysD2019} to provide additional perpendicular anisotropy, but this has negative consequences for device resistance and is incompatible with spin-orbit torque switching \cite{PrenatIEEE2016} which has a speed and durability advantages for certain applications. 

While useful for research purposes the prototypical MTJ has a number of deficiencies as a practical MRAM device, requiring high annealing temperatures to crystallize the CoFeB layers and having a large shift in the threshold switching current for parallel and anti-parallel orientations of the free layer due to the stray magnetic field originating from the reference layer\cite{BapnaAPL2016}. Practically this is compensated by adding a pinned layer (PL) which is magnetically stable and coupling this layer antiferromagnetically to the reference layer, forming a synthetic antiferromagnet structure, or SAF. The antiferromagnetic coupling between the PL and RL is engineered by using a thin metallic layer of Ir or Ru which mediates the RKKY exchange interaction across the layers \cite{ParkinPRL1991}. The thickness of the PL can be adjusted to reduce the stray magnetic field at the free layer position and therefore reduce the asymmetry in the threshold STT switching current. A simplified stack structure with a SAF included is shown in Fig.~\ref{fig:stacks}(b) with the addition of bottom pinned layer (PL) and exchange coupling layer. Here we have assumed that the PL is stabilized partially by the addition of a bottom MgO layer to provide high anisotropy and also by exchange coupling to the reference layer. 

The addition of the SAF mitigates the problem with the stray magnetic field originating from the reference layer but is somewhat inflexible, requiring precise fabrication of magnetic layers with atomic level precision. Practical devices therefore typically use thicker magnetic layers more resistant to small fabrication divergences and also provide more flexibility in materials choices, including heavy metal doping to improve crystallization and diffusion during manufacture. For thicker layers the interfacial anisotropy from the MgO tunnel barrier is no longer sufficient to sustain perpendicular anisotropy, and so typically a CoPt alloy or multilayer is used to provide additional perpendicular anisotropy for the thicker layers, shown in Fig.~\ref{fig:stacks}(c). Some devices optionally include an antiferromagnetic layer beneath the pinned layer to provide an unconditionally stable exchange bias field to ensure long-term stability of the pinned layer magnetization. Typically the antiferromagnet is IrMn or PtMn due to the high N\'eel temperature and large magnetic anisotropy \cite{Vallejo-FernandezAPL2010,Vallejo-FernandezAPL2007,SzunyoghPRB2009,JenkinsJAP2018}. While the bulk magnetization of an antiferromagnet is essentially zero, at the nanoscale atomic lattice defects and the non-collinear nature of the antiferromagnetic spins may lead to a magnetic stray field not usually accounted for in MRAM device designs.

\section{Atomistic dipole fields}
Most studies of stray magnetic fields utilise either classical Maxwellian magnetostatics for simple geometrical shapes \cite{Jackson1975}, or numerical micromagnetics where the magnetic vector potential is considered in the continuum limit. While such approaches are suitable for large scale devices, the exceptionally thin films and device sizes less than 50 nm needed for MRAM approach the limits of applicability of the continuum approximation. At the electronic level the spin polarised electron density is a continuous property of a magnetic material, but with a strong spatial dependence and localised in the vicinity of the atomic nuclei even for classically itinerant magnets such as Fe and Co \cite{SchwarzJPhyF1984}. Where the moments are well-localised the dipole-dipole approximation\cite{Jackson1975} is often employed which considers each atom as a point source of magnetic field and is a good approximation when considering most magnetic materials. Considering an atom at any point in space $i$ experiences a dipole (induction) field $\mathbf{B}_i$ from all nearby magnetic dipole moments $j$, the magnetic field is given by
\begin{equation}
\mathbf{B}_i = \frac{\mu_0}{4\pi} \sum_{j} \left[ \frac{3\mathbf{\hat{r}}_{ij} (\mathbf{\hat{r}}_{ij} \cdot \mathbf{m}_j) - \mathbf{m}_j}{|\mathbf{r}_{ij}|^3} \right]
\label{eq:atomic-dipole}
\end{equation}
\noindent where $\mathbf{r}_{ij}$ is the distance between point $i$ and the magnetic moments at point $j$, $\mathbf{\hat{r}}_{ij}$ is a unit vector from site $i$ to $j$, $\mathbf{m}_j$ is the magnetic moment at site $j$, and $\mu_0 := 4 \pi \times 10^{-7}$ H/m.

In the above definition we explicitly exclude the self-term acting within each dipole, since this field always opposes the dipole magnetic moment and has no effect on the dynamics of local moments. As noted by Kittel \cite{Kittel2004}, the dipole field at the centre of a spherical lattice of dipoles is zero at the centre, which is different from the Maxwellian field of $\mathbf{H} = -\mathbf{M}/3$ found in micromagnetic calculations. Classically this is resolved by invocation of a Lorentz sphere which provides the apparently absent demagnetizing field \cite{Kittel2004}. However, with modern computational approaches we can compute the dipole field of a large (100 nm) finite sphere exactly which naturally agrees with the analytical limit that the dipole field at the centre of a sphere is zero. The origin of this discrepancy is likely the self term for point dipoles \cite{Jackson1975} though the resolution of a disagreement between dipole and Maxwellian fields is beyond the scope of the present article. It is important however to state the difference in the two approaches and for an infinite thin film the local demagnetizing field computed from the dipole-dipole approximation in Eq.~\ref{eq:atomic-dipole} is $\mathbf{H} = -2\mathbf{M}/3$ rather than $\mathbf{H} = -\mathbf{M}$. Outside the magnetic material the computed magnetic field is of course identical between the Maxwellian micromagnetic and dipole-dipole approach. In the following analysis we neglect the self-field within the magnetic material and include only the \textit{free} magnetic induction arising from the dipoles, i.e. $\mathbf{B} := \mu_0 \mathbf{H}$ where $\mathbf{H}$ is the dipole-dipole field. 

The dipole-dipole interaction decays proportional to $1/|\mathbf{r}_{ij}|^3$ and so the long-range nature of the dipole-dipole interaction requires significant computational power. For a system of $N$ atoms each dipole is interacting with $N-1$ dioples and, thus, an atomistic calculation would lead to a computational complexity proportional to $N(N-1) \sim N^2$. To make such calculations feasible we have implemented a massively parallel and scalable calculation of the atomistic dipole-dipole field within the \textsc{vampire} code~\cite{EvansJPCM2014,vampire}. For the parallel atomistic dipole-dipole solver we first collate atomic positions, moments and spin directions from each processor in the calculation onto \new{the main processor}, since the standard parallelisation in the \textsc{vampire} code~\cite{EvansJPCM2014,vampire} uses a parallel geometric decomposition where the moments are distributed \new{among all} the processors~\cite{vampire}. \new{The positions, moments and spins are then broadcast to all processors so that every processor has a complete copy of the system. Although this is expensive in memory, for moderately sized systems of a million spins this is only tens of megabytes (MB) per processor. The advantage of this approach is that each processor now has access to the complete set of spin, moment and position data and is able to compute the dipole-dipole field calculation for any spin $i$.} The fields for local moments on each processor are then computed by considering all other dipole moments in a simple brute force approach, computing Eq.~\ref{eq:atomic-dipole} directly for every other magnetic dipole moment in the system. We split this calculation into two separate processing loops for $i < j$ and for $i>j$ to avoid the redundant check that $i \ne j$ within the main computation loop to improve performance. Our parallel implementation is highly scalable with a computational cost of approximately $N^2/N_p$ where $N_p$ is the number of processors used for the computation. Typically $N^2 >> N_p$ leading to near ideal scaling for the computational complexity of this part of the calculation. This allows the calculation of direct dipole-dipole interactions for systems of 1M dipoles on a few tens of processor cores in a few minutes.

\section{Results}
Having defined the different basic kinds of MRAM device structures, we now consider the title problem: the strength and anisotropy of the magnetic stray field from the different magnetic layers of a device. We consider an idealistic MRAM device uniformly patterned into a 25 nm diameter cylinder.

\begin{figure}[tb]
\centering
\includegraphics[width=8.5cm]{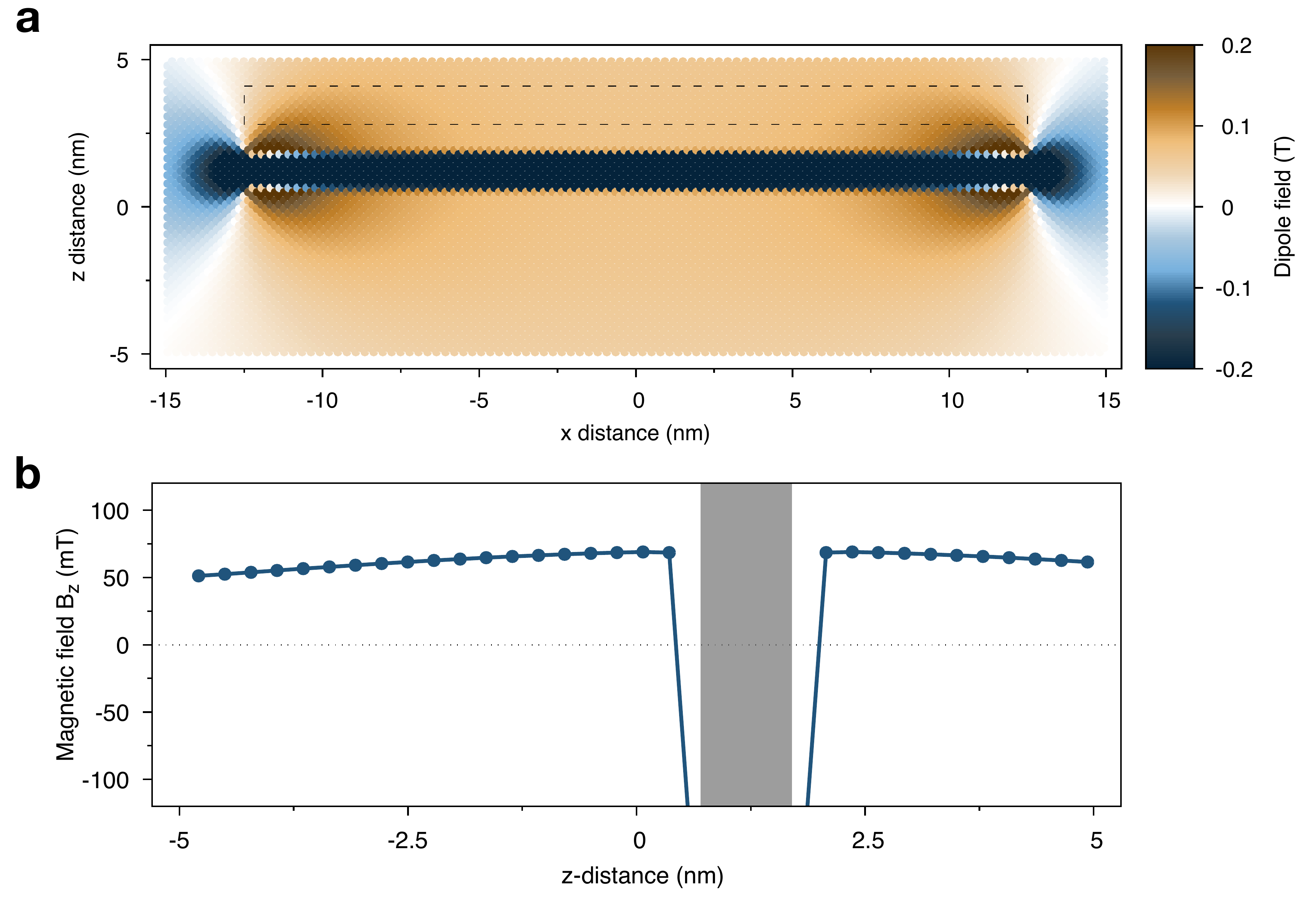}
\caption{Computed stray magnetic field emanating from a single reference layer 1 nm thick in the $x-z$ plane (a). The legend is capped at maximum fields of $\pm$ 200 mT to better show weaker fields in the vicinity of the free layer shown by the dashed line. The calculated axis field along the line $x=y=0$ is shown in (b) showing a slow decay away from the reference layer\new{, indicated by the shaded area}. Color Online.}
\label{fig:single}
\end{figure}

\subsection{Prototypical bilayer MTJ}
Let us first consider the prototypical bilayer MTJ with reference and free layers, \new{shown schematically in Fig.~\ref{fig:stacks}(a)}. Here the reference layer is fixed and emits a stray magnetic field aligned with the magnetization of the layer. \new{As noted earlier, we compute only the magnetic field in free space and ignore the Maxwellian self-term contribution within the magnetic material. Both magnetic layers are assumed to be CoFeB with a saturation magnetization of $\Ms = 1.35$ MA/m.  We omit the free layer from all our calculations as we wish to study the stray field emanating directly from the reference (and later pinned) layers.} The computed strength and $z$-component of the magnetic field emanating in the vicinity of the reference layer (uniformly magnetizated along the $+z$-direction) is shown in Fig.~\ref{fig:single}(a). The net magnetic field at each point is computed directly from the full 3D problem of atomic source dipoles given by Eq.~\ref{eq:atomic-dipole}. Within the reference layer the dipole field opposes the magnetization and is much larger than the field outside the device. The colour scale is saturated at $\pm 0.2$ to better highlight the structure of the stray field outside the magnetic layer, where larger fields are displayed with the saturated colour intensity. The position of the free layer above the reference layer is indicated by the dashed line. As expected for any free ferromagnet, the stray field is emitted parallel to the magnetization leading to a net positive bias field of around +65 mT at the position of the free layer. This naturally leads to a bias of the minor hysteresis loop \cite{DevolderJPhysD2019,MeoSciRep2017,IkedaNatMat2010} and a similar shift of the threshold current for spin transfer torque switching and is undesirable for device operation. The field strength along the centre axis of the nanodisk is shown in Fig.~\ref{fig:single}(b) showing a slow decay of the field strength moving along the $z$-axis away from the magnetic layers. At the dot edges the stray field is highly non-linear due to the need for flux closure and leads to large magnetic fields in excess of 100 mT at the free layer edges. For larger device diameters the edge effect is less important as the low flux region in the centre of the device dominates the average field, but for small diameters these large fields will become much more dominant. 

\subsection{Defects in bilayer MTJs}
\begin{figure}[htb]
\centering
\includegraphics[width=8.5cm]{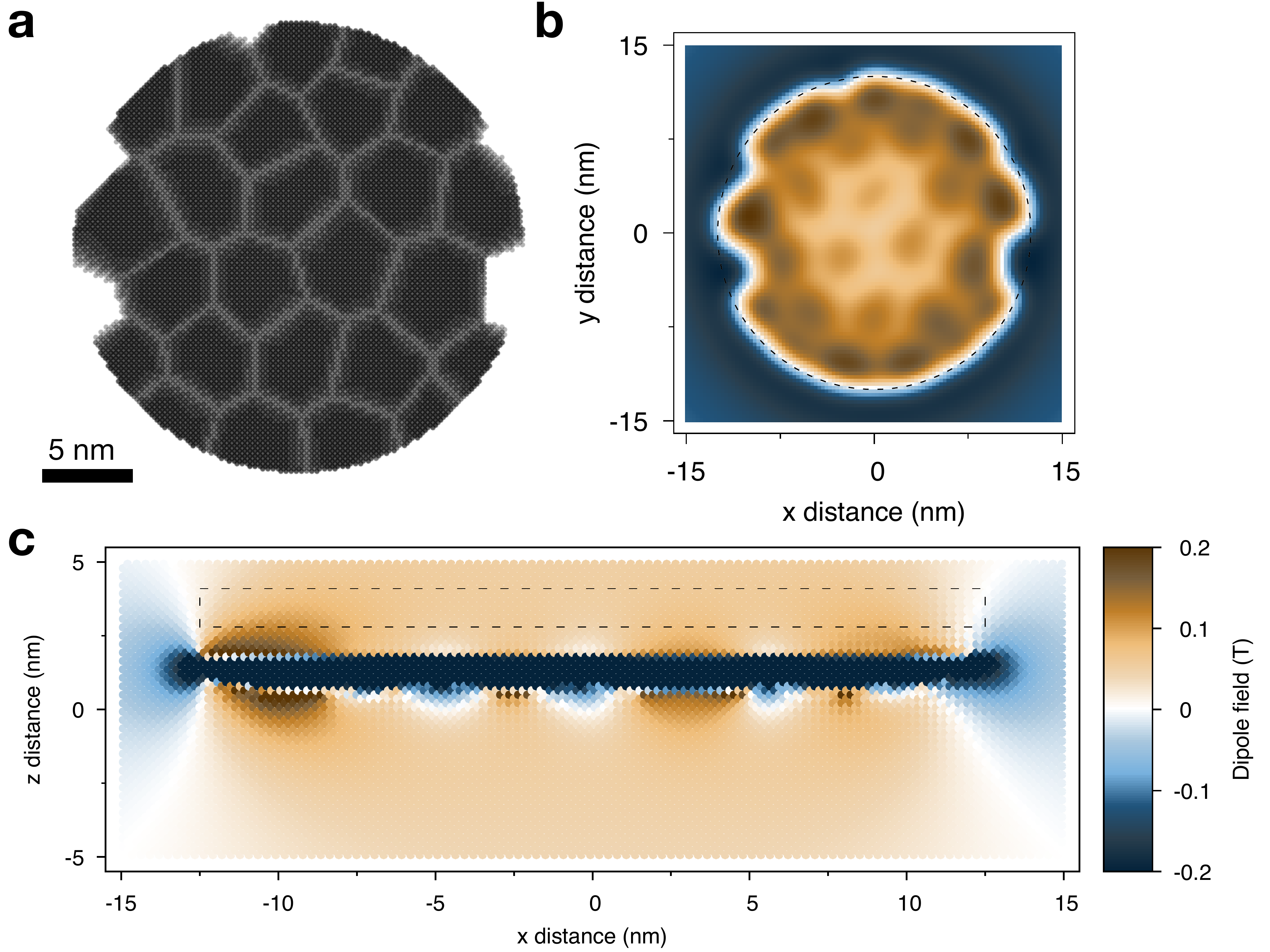}
\caption{Computed stray fields from a polygranular reference layer. (a) Visualization of the polygranular structure of the layer and edge defects arising from the patterning process. (b) Top-down view of the stray field computed at the centre of the free layer showing an imprint of the polygranular structure in the stray magnetic field. (c) Side view of the computed stray field for the polygranular structure, showing non-linearities and flux closure close to the interface. Color Online.}
\label{fig:poly}
\end{figure}

An important consideration for nanoscale devices is the role of defects arising due to deposition, annealing and patterning of the devices. The diversity of such effects is an expansive topic and we are only beginning to be able to address their relative importance to device operation, however we are able to consider the likely polygranular nature of annealed CoFeB/MgO. This arises due to the polygranular nature of the thin MgO layer \cite{WangNL2016} which is imparted to the amorphous CoFeB layers during annealing and crystallization \cite{PellegrenIEEE2015}. We model this by considering a polygranular structure to the device generated using a voronoi tessellation. An example structure is shown in Fig.~\ref{fig:poly}(a) which has been patterned into a 25 nm diameter cylinder. Additionally some of the edge grains have been removed during the patterning process to simulate patterning defects which may occur at such sizes. The role of a polygranular structure on the overall magnetic properties and switching dynamics will be the subject of a future study\cite{ElliotGranular2019}, but here we consider the stray field from a single polygranular reference layer in a simple bilayer MTJ topology, shown in Fig.~\ref{fig:poly}(b,c). The stray field at the free layer position visibly adopts the underlying structure of the polygranular reference layer shown in the top-down view in Fig.~\ref{fig:poly}(b), in particular the edge defects which visibly affects the non-linear field at the device edges. The side view in Fig.~\ref{fig:poly}(c) shows a similar average field profile to the single continuous layer in Fig.~\ref{fig:single}(a), but flux closure and non-uniformities are clearly visible near the layer interface. Collectively even simple defects add additional complexity when considering the stray field in devices and are of course random in nature. This will naturally impact the consistency of device operation when considering gigabit device arrays and may be an additional factor to consider in device manufacture, particularly at smaller process nodes.

\new{In the case of prototypical MTJs with and without defects} the non-uniform magnetic fields contribute to three effects. The first is a large asymmetry of the hysteresis loop, seen as a bias field shift of the loop to one side depending on the magnetic orientation of the reference layer \cite{BapnaAPL2016}. The second is a different threshold switching current considering the parallel to anti-parallel (P $\rightarrow$ AP) orientations of the reference and free layers, and anti-parallel to parallel (AP $\rightarrow$ P). This second effect has the same physical effect as the first, with a simple bias field. This adds an effective magnetic anisotropy to one of the two configurations (e.g. P), and reduces the effective anisotropy for the opposite orientation (e.g. AP). This therefore increases the current required to initiate STT switching for the orientation of the larger effective anisotropy configuration, and provides a comparable reduction in the threshold current for the lower anisotropy configuration. The third effect of the non-uniform magnetic fields is to influence the nature of the reversal mechanism. The reversal of nanoscale dots is usually assumed to be coherent \cite{OFlynnJAP2013}, while energy barrier simulations \cite{OFlynnJAP2013}, room-temperature atomistic simulations \cite{MeoSciRep2017} and experimental measurements \cite{BapnaAPL2016} find that the reversal is edge nucleated due to thermal fluctuations. Stray field non-uniformities at the dot edges will contribute an additional preference for nucleated reversal, though at room temperature the reversal mechanism of dots below the single domain limit $\sim 20$ nm in  diameter is already dominated by thermal effects and are superparamagnetic \cite{MeoSciRep2017}. Importantly, the strength of the non-uniform stray field edge effects is probably of secondary importance to the reversal mechanism compared to thermal fluctuations, since these are dominant for such thin films and small devices \cite{MeoSciRep2017}. In contrast, the average stray field at the free layer will lead to a macroscopic asymmetry of the hysteresis loop and switching current. Compensating these stray-field effects is essential for reliable device operation and we now consider the addition of an antiferromagnetically coupled pinned layer to compensate the stray field from the reference layer at the location of the free layer.

\begin{figure*}[htb]
\centering
\includegraphics[width=14.5cm]{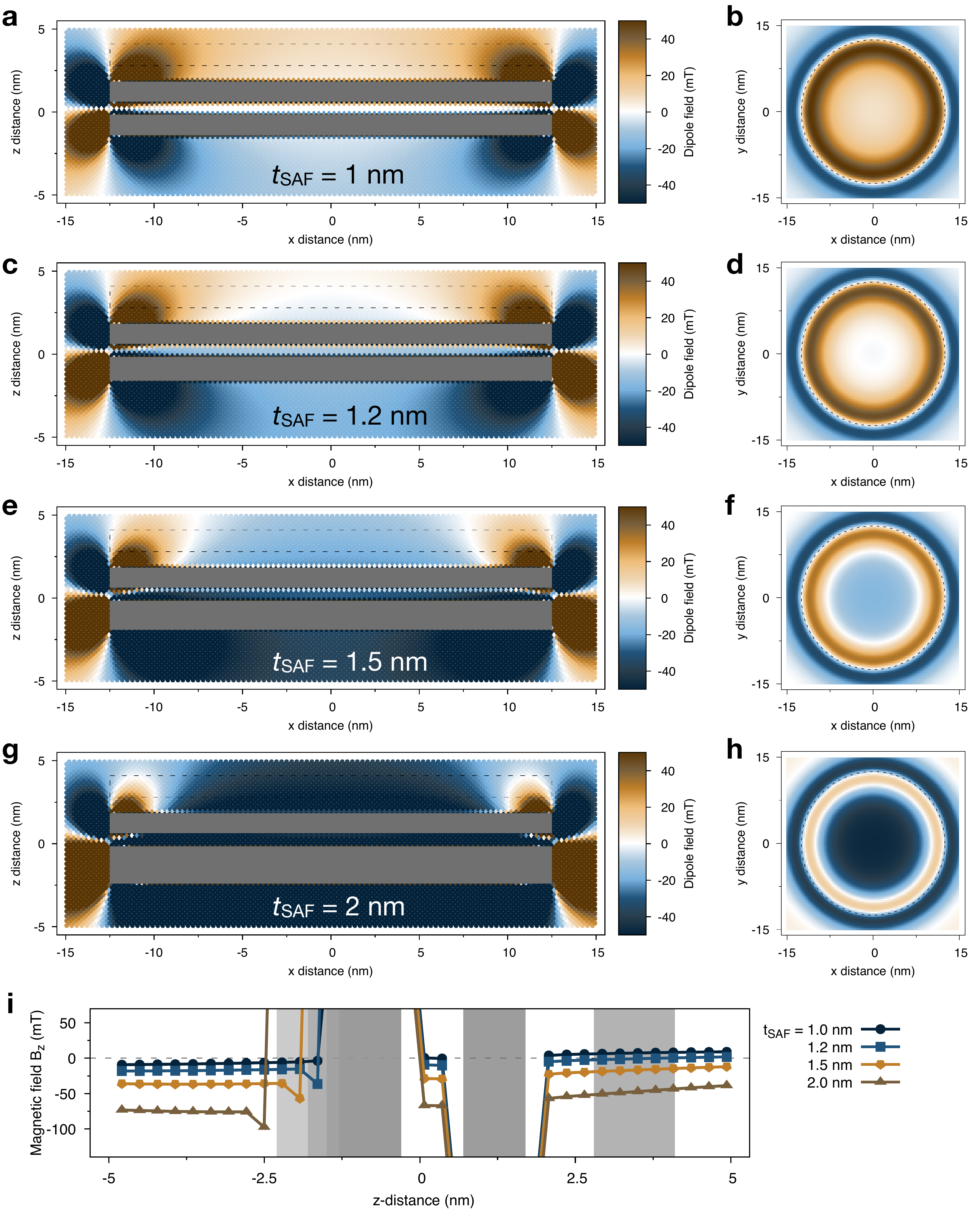}
\caption{Computed dipole fields in the planes $y = 0$ and $z = 3 nm$ (at the centre of the free layer) for pinned layer thicknesses of 1 nm (a,b), 1.2 nm (c,d), 1.5 nm (e,f) and 2 nm (g,h). The colour indicates the magnitude and direction of the z-component of the net dipole field at each point. The colour key saturates at $\pm 50$ mT to focus on the low field data. The position of the free layer is indicated by the dashed line assuming a spacing of 1 nm of MgO above the central reference layer. \new{An axial line profile for each pinned layer thickness is shown in panel (i) showing the net cancellation of the field at the free layer position.} The magnetic layers are blocked out in grey to clearly show the stray field regions. (Colour Online).}
\label{fig:demag}
\end{figure*}

\subsection{MTJ with SAF geometry}

Here we consider a simplified structure based on the prototypical CoFeB/MgO/CoFeB MTJ nanodot structure consisting of a 1 nm thick reference layer, 1.3 nm thick free layer (not included) and variable thickness pinned layer, $t_{\mathrm{P}}$ with a cylindrical device diameter of 25 nm\new{, shown schematically in Fig.~\ref{fig:stacks}(b)}. Collectively the pinned and reference layers form the \saf. As before the magnetization of the reference layer is set along the $+z$ direction while the pinned layer set along the $-z$ direction to attempt to reduce the strength of the stray field emanating from the reference layer at the free layer position. Figure.~\ref{fig:demag} shows slices through the computed z-component of the stray field for different thicknesses of the bottom pinned layer. For the symmetric case where both the pinned and reference layers are 1 nm thick in Fig.~\ref{fig:demag}(a), the stray field from the \new{pinned and reference} layers is \new{anti-symmetric and exactly zero between the two layers. When the free layer is included the symmetry is naturally broken, but here we are only interested in the net field at the free layer location.} Considering the central axis of the MTJ ($x = 0$), the stray field only approaches zero between the two layers, with a low field with opposite polarity as one moves away from the structure. The field along the $x=y=0$ axis  \new{shown in Fig.~\ref{fig:demag}(i)} is weaker than for the isolated case but the relative proximity of the two oppositely magnetized layers leaves a stray field of approximately 20 mT at the free layer location. A clear feature of the nanoscale device including a SAF is the persistent large edge field necessary for flux closure as with the simple bilayer device. This field is highly non-linear within the space for the free layer indicated by the dashed line. As expected this field is symmetric around the circumference of the dot as shown in Fig.~\ref{fig:demag}(b). Due to the cylindrical nature of the device the \new{fringing} field makes a large contribution to the areal average stray dipole field at the free layer position. 

Expanding the pinned layer thickness to 1.2 nm increases the moment and therefore decreases the field at the centre of the free layer position to less than 10 mT, accounting for the closer proximity of the reference layer as shown in Fig.~\ref{fig:demag}(c,d). The width of the high-field edge region shown in Fig.~\ref{fig:demag}(d) is reduced compared to the single layer but still makes up a large fraction of the average areal field at the centre of the free layer location. While the field along the central axis is significantly reduced, the edge effects still remain with a large fringing field at the device edges. This fundamentally compromises the role of the SAF in compensating the average field and demonstrates the importance of edge field effects. 

Further increasing the pinned layer thickness to 1.5 nm as shown in Fig.~\ref{fig:demag}(e,f) now overcompensates the stray field in the axial region of the free layer with a small negative field. However, the fringing field in the edge region is both narrowed and weaker compared with the 1.2 nm thick layer. The overcompensating field in the centre of the free layer now balances the fringing field so that the \textit{average} field across the device approaches zero, but now with competing dipole field contributions at the centre and edge of the free layer. This reduces the strength of the edge field which contributes to the edge nucleation reversal mode and therefore may favour a more coherent reversal mechanism.

In Fig.~\ref{fig:demag}(g,h) a large pinned layer thickness of 2 nm is included. The stray field from the pinned layer now dominates the reference layer, with large negative fields at the free layer position along the $x=y=0$ axis. The edge effects are much weaker than for thinner pinned layers but clearly the compensating role of the pinned layer is no longer working. However, some engineered bias field on the free layer may be beneficial for STT switching. For STT switching there is a natural imbalance in the P $\rightarrow$ AP and AP $\rightarrow$ P switching thresholds due to the different origin of the spin torque. For the AP $\rightarrow$ P case the spin \textit{transmitted} through the reference layer provides a torque on the free layer causing it to align with the reference layer. For the P $\rightarrow$ AP switching case the smaller \textit{reflected} spin current is responsible for generating a torque on the free layer, therefore requiring a larger current to switch to the AP configuration. These effects are partially compensated by the low and high device resistance in the P and AP states respectively which naturally increases the current flow in the P configuration. However, a weak energetic preference for the AP configuration would reduce the threshold current for P $\rightarrow$ AP switching and may be advantageous for device operation. While not sensible for a traditional SAF, an overcompensating pinned layer may be advantageous for STT-MRAM devices.

\subsection{Stray field from an antiferromagnet}
Finally we consider the stray field from a nanoscale antiferromagnet, used as an exchange biasing layer to make the pinned layer unconditionally stable. Practically this is important in terms of the resilience of MRAM devices to large external magnetic fields. If the chip is exposed to a sufficiently large magnetic field to reverse the pinned layer, then for a uniaxial pinned layer the device would no longer function. In contrast, the unidirectional nature of the exchange biased pinned layer means that the data would likely be erased but the device would still function once the field is removed. Being antiferromagnetic, one usually assumes that the stray field emanating from it is zero, since there is no net magnetic moment. However, at the atomic scale the magnetic moments are quite large and so close to the layer one might expect some small stray fields. In addition, edge and interface effects can lead to a small net moment in the antiferromagnet, which is of course required for exchange bias to work. 

To assess this we model a 5 nm slab of L$1_2$-ordered \IrMn using an atomistic spin model\cite{JenkinsJAP2018,JenkinsArXiV2019}. The energetics of the system are described by the spin Hamiltonian:

\begin{equation}
\Ham = -\sum_{i<j} \Jij \sS_i \cdot \sS_j - \frac{k_N}{2} \sum_{i \neq j}^z (\mathbf{S}_i \cdot  \mathbf{e}_{ij})^2 
\label{eq:hamiltonian}
\end{equation}

where $\sS_i$ is a unit vector of the spin direction on a Mn site $i$, $k_N = -4.22 \times 10^{-22}$ is the \Neel pair anisotropy and $\mathbf{e}_{ij}$ is a unit position vector from site $i$ to site $j$, $z$ is the number of nearest neighbours and \Jij is the exchange interaction. The exchange interactions were limited to nearest ($\Jnn = -6.4 \times 10^{-21}$ J/link) and next nearest ($\Jnnn = 5.1 \times 10^{-21}$ J/link) neighbours~\cite{JenkinsJAP2018}. The system is initialised with a random spin configuration and then zero-field cooled using an adaptive Monte Carlo\cite{AlzateCardonaJPCM2019} to form a single domain ground-state spin structure with triangular (T1) symmetry \cite{Kohn,TomenoJAP1999,SzunyoghPRB2009}.

\begin{figure}[tb]
\centering
\includegraphics[width=8.5cm]{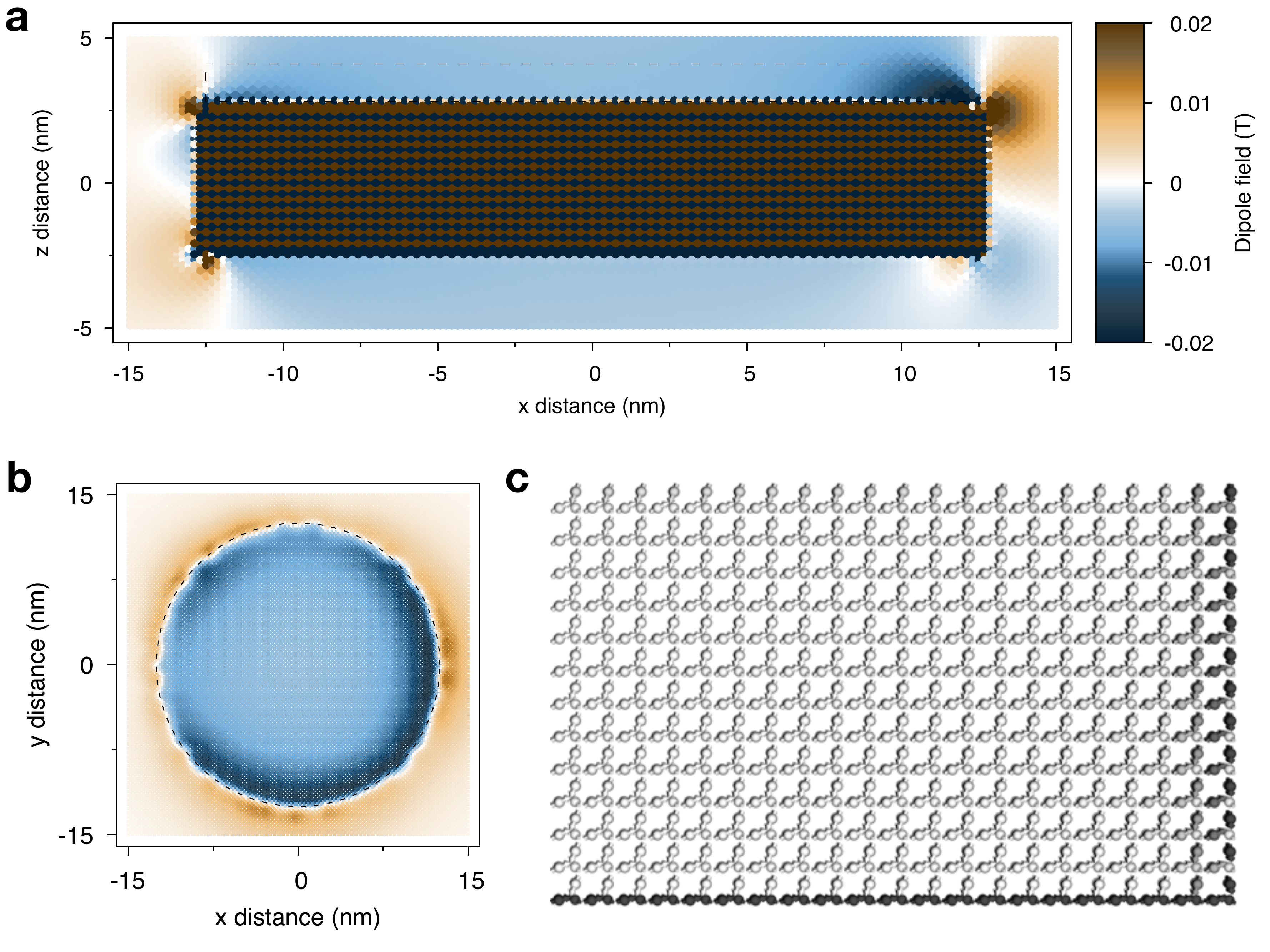}
\caption{Computed stray fields from a 5 nm thick IrMn exchange biasing layer. (a) Side view of the computed stray field for the IrMn layer, showing a non-zero stray field and edge effects. (b) Top-down view of the stray field computed at the centre of an adjacent magnetic layer. (c) Computed spin configuration in the antiferromagnet showing the non-collinear nature of IrMn. The contrast indicates the degree of local spin deviation from the collinear state, saturating at 1\% $(\mathbf{S}_i \cdot \mathbf{n})$ where $\mathbf{n}$ is the sublattice magnetization. Color Online.}
\label{fig:irmn}
\end{figure}

The stray field is computed as above for ferromagnetic layers using the direct dipole-dipole interaction using Eq.~\ref{eq:atomic-dipole} and plotted in Fig.~\ref{fig:irmn}. Here we consider the stray field generated within a ferromagnet placed in direct contact with the antiferromagnetic layer\new{, but as for previous calculations the stray field from the ferromagnet is not calculated}. Considering first the cross-section of the computed field in Fig.~\ref{fig:irmn}(a) it is clear that within the antiferromagnet the dipole fields are quite strong, and likely add additional magnetic anisotropy. What is most surprising is the non-zero stray field emanating from the bulk of the antiferromagnet which is approximately 5 mT along the $-z$-direction along the central axis of the disk. While the strength of the field is an order of magnitude weaker than that of ferromagnetic layers, its non-zero nature is in direct contrast to conventional wisdom regarding antiferromagnets. The edge field is similarly weaker than in ferromagnetic layers and also exhibits some rotational asymmetry as shown in Fig.~\ref{fig:irmn}(b). The asymmetry in the edge field arises due to different edge crystal terminations and therefore a slight imbalance in the number of moments in each magnetic sublattice when considering different surface contributions. This also explains the observation of a net stray field from the antiferromagnet, by considering net magnetic moments on the surface of the system arising from the imbalance of magnetic sublattices. These net moments then form a surface contribution to the dipole field which then exhibits a macroscopic stray field behaviour. 

\new{To illustrate the surface effects of the termination we show a slice near $(y = 0)$ of the atomic spin structure in Fig.~\ref{fig:irmn}(c) represented by arrows. The contrast shows the deviation of the local spin direction from the bulk sublattice magnetization, with black arrows representing a 1\% deviation from $(\mathbf{S}_i \cdot \mathbf{n})$ where $\mathbf{n}$ is the sublattice magnetization. White arrows indicate 0\% deviation from the collinear state. The sublattice ordering over the whole dot is greater than 99\% confirming the single domain nature of the antiferromagnet, and the small reduction in order is due to surface spin canting resulting from the loss of coordination at the surface and therefore inducing a small local spin canting. The data show the existence of a single plane of collinear atoms at the top surface which is likely the source of the small stray field in the vicinity of the surface. It is clear from the data that there is a weak surface canting of spins at the side walls of the nanodot but these are visually symmetric suggesting that they are not the direct origin of the asymmetry in the fringing field considering the $\pm x$ sides of the nanodot. Therefore the origin of the asymmetry must be the complex interplay between the surface crystal faceting and the sublattice magnetization, where the dominance of one particular sublattice at a particular surface leads to a different local stray field. This view is supported by the data in Fig.~\ref{fig:irmn}(b) showing a continuous variation of the stray fringing field.} 

The specific stray field from an exchange biasing antiferromagnetic layer is likely to be specific to the antiferromagnetic spin structure, crystal termination and defects and therefore hard to deterministically account for in device design. The stray field generated from and antiferromagnet is therefore an additional source of dispersion of single device properties that could negatively impact on consistency of device properties when considering the thermal stability.

\section{Conclusions}
In conclusion, we have studied the stray fields emanating from nanoscale layers in magnetic tunnel junctions using an atomistic dipole-dipole approach. We have found that edge effects make a significant contribution to the effective dipole field at the free layer position in agreement with previous calculations\cite{DevolderJPhysD2019}. Considering a range of thicknesses for a compensating pinned later in the synthetic antiferromagnetic structure we find incomplete cancellation of the stray field from the reference layer with persistent non-linear fields at the dot edges. A slightly over-compensated field may have some benefits in compensating for asymmetry in the threshold switching current considering spin transfer torque switching for P $\rightarrow$ AP and AP $\rightarrow$ P configurations. We have considered the role of a defected granular structure on the stray field from a single ferromagnetic layer and find that patterning defects have a strong influence on the edge stray field and the granular structure is imparted to the free layer with a non-uniform field. The stray field in close proximity to the grains exhibits flux closure which may be important considering very thin layers magnetic layers in close proximity. Finally we have considered the stray field from an antiferromagnetic layer and have found that the stray field is non-zero at the nanoscale due to imperfect cancellation of the sublattice magnetization at the surfaces. This stray field makes an additional contribution to the thermal stability of the pinned layer which leads to a \new{natural distribution of device properties}.

While we have studied only a fixed size device of 25 nm due to computational limitations, the edge effects are quite general, and will give a smaller contribution to the average field in the free layer for larger devices and more significant for smaller devices. The strength of the edge field suggests that it may be beneficial to pattern the free layer with smaller dimensions than the reference layer so that it is contained entirely within the uniform region of the stray field, as previously proposed by \cite{BapnaAPL2016} \textit{et al}. The effectiveness of the SAF structure is also more challenging at the nanoscale due to these significant edge effects, and so more complex designs could be considered, with a thicker circumferential (ring-like) compensating pinned layer to counteract the edge effects. \new{For smaller devices approaching 5 nm in diameter the fringing field will be dominant with no uniform axial component, making field cancellation using a SAF particularly difficult. This presents additional challenges for the manufacturing of such small devices and may require a different geometry such as a continuous granular pinned layer spanning multiple devices to ensure uniformity of stray fields.} 

Defects present a particular challenge considering dipole fields, since the film morphology can influence the specific characteristics at the nanoscale. In particular orange-peel coupling effects can become important \cite{TsaiJAP2014} and even percolated exchange coupling\cite{SokalskiAPL2009,EvansJPhysD2014,EllisAPL2017}. Future devices utilizing shape anisotropy to enhance thermal stability for sub-20nm lateral dimensions \cite{WatanabeNatComms2018,PerrissinRSC2018,PerrissinJPhysD2019} rely on a full understanding of dipole interactions at the nanoscale, and so similar atomistic calculation methods presented here can be used to model the role of different physical defects on the effective thermal stability and in particular their switching dynamics.

\section{acknowledgements}
We gratefully acknowledge the provision of computer time made available on the \textsc{viking} cluster, a high performance compute facility provided by the University of York. This work used  code enhancements implemented and funded under the ARCHER embedded CSE programme (eCSE1307) supported by the ARCHER UK National Supercomputing Service (http://www.archer.ac.uk). S.A.M. acknowledges support from US NSF grant ECCS-1709845. The authors thank Rob Carpenter and Graham Bowden for helpful discussions.

\bibliography{local}

\end{document}